\documentclass[prb,twocolumn,showpacs,showkeys,preprintnumbers,amsmath,amssymb]{revtex4}
\usepackage{graphicx} % Include figure files
\usepackage{dcolumn}  % Align table columns on decimal point
\usepackage{bm}       % bold math

\begin{document}
\flushbottom

\preprint{doping-graphene}

\title{Barrierless procedure for substitutionally doping graphene
sheets with boron
atoms: \textit{ab initio} calculations
}
\author{Renato B. Pontes, A. Fazzio and Gustavo M. Dalpian}

\affiliation{
Centro de Ci\^encias Naturais e Humanas, Universidade Federal do ABC, Santo Andr\'e, SP, Brazil \\ Instituto de F\'\i sica, Universidade de S\~ao Paulo, CP 66318, 05315-970, S\~ao Paulo, SP, Brazil}
\date{\today}

\begin{abstract}
Using ab initio methods, we propose a simple and effective way
to substitutionally dope graphene sheets with Boron. The method consists of
selectively exposing each side of the graphene sheet to different elements. 
We first expose one side of the membrane to Boron,
while the other side is exposed to Nitrogen.
Proceeding this way, the B atoms will be spontaneously incorporated into the graphene membrane, 
without any activation barrier. In a second step, the system should be exposed to a H-rich environment,
that will remove the CN radical from the layer and form HCN, leading to a perfect substitutional doping.
\end{abstract}

%72.80.Rj Fullerenes and related materials 
%74.62.Dh Effects of crystal defects, doping and substitution
%61.72.U- Doping and impurity implantation 
%61.72.-y Defects and impurities in crystals; microstructure
\maketitle

%% begin paper
Graphene recently became one of the most studied materials in the literature due to its potential applications for novel devices, 
hydrogen storage and due to its unique physical and chemical properties\cite{1,2,3,4,5}. Nowadays, simple graphene-based devices are already feasible, showing that the initial concepts proposed for graphene can be manufactured\cite{6}. On the other hand, in order for functional devices to be developed, a more precise ability to manipulate graphene has to be reached. In order to do that, new functionalization processes have to be developed. A very effective way to achieve this functionalization level is through doping. As most materials have to be doped to build functional devices with them, it is reasonable to argue that graphene also should be doped in order to build functional devices with it. 

Doping of nanostructures is known to be a very difficult task\cite{7,8}. This difficulty comes from
a combination of energetic and kinetic effects. 
 The same happens for graphene, with few successful cases of substitutional doping. 
It is known that defects can be created by ion-irradiation\cite{9} and impurities can be incorporated
 by pyrolitic\cite{10} processes. 
High-Temperture pyrolitic processes, for example, are adequate for applications such as hydrogen storage, but are too dirty for electronic purposes. A better-controlled method is needed for that. Doping can also be induced by the adsorption of molecules on top of graphene\cite{12}, but precise atom substitution by foreign impurities is still a challenge. 
Here we propose a procedure to substitutionally dope graphene sheets, ribbons and possibly carbon nanotubes.
The process described here  does not have any activation barrier, i.e., it occurs spontaneously once the reagents are placed
in different sides of graphene.
This system 
is a potential candidate for high-performance Hydrogen storage material\cite{10,11}, nanosensors and spin filter devices\cite{13}. 

Our methodology has two steps: ({\it i}) in the first step, one side of the graphene membrane should be exposed to a 
B-rich environment, while in the other side a N-rich environment should be created. 
This combination will make the system energetically unstable, and  
the consequence is that a C atom will be pushed out from the sheet, and on its place a B atom will be incorporated;
({\it ii}) the second step is the removal of the C-N radical that is bonded to the B atom. In order to do that, the system should be inserted into an H-rich medium, or an acid medium. 
To simulate this, we insert a H atom in the N-rich side of the membrane. The H atom will bind to the C-N, 
removing it from the surface and forming HCN, completing the B-doping process. Once again, this is a barrierless process.

Although complex from an experimental point of view, according to recent results, our approach should be feasible. Recent measurements\cite{14} show that macroscopic graphene sheets can be created and precisely placed on top of a support grid for transimission electron microscopy. This is a possible experimental apparatus necessary for the creation of the membrane in order for our doping process to be realized. This support could be used as
a separator for the two environments. Other recent experiments show that it is possible to apply a pressure difference
across a graphene membrane, and that it should be impermeable to standard gases,  providing {\it  a unique separation barrier between two distinct regions that is only one atom thick}\cite{bunch}. Our model also shows that, besides separating two distinct regions, depending on the configuration of each medium, atom substitution should occur.

To study this process, we performed spin-polarized total energy calculations based on the density functional 
theory\cite{15} within the generalized gradient approximation\cite{16}. We also used norm conserving 
pseudopotentials\cite{17}, as implemented in the SIESTA code\cite{18}. We employ a {\it double zeta} basis
 function with polarization orbitals \cite{emilio}, a confining energy shift of $0.06\, eV$  and a mesh cutoff energy of $250\,Ry$
 for the grid integration. The Brillouin zone was sampled by using a Monkorst-Pack \cite{pack} scheme with a ($15 \times 1\times 15$)
 k-point sampling. Smaller k-point samplings leads to an incorrect description of the density of states of graphene, 
leading to possibly incorrect results. Periodic boundary conditions were used in the simulations, 
and the graphene sheet was described by a ($3 \times 4$) supercell with 48 carbon atoms. A vacuum region of 15 \AA\, was 
used in order to separate the layer and its images in the direction perpendicular to the plane.
 All atoms of the system were fully relaxed within a force convergence criterion of 25 meV/\AA.

\begin{figure}[ht]
\includegraphics[angle=0,width=9.cm ]{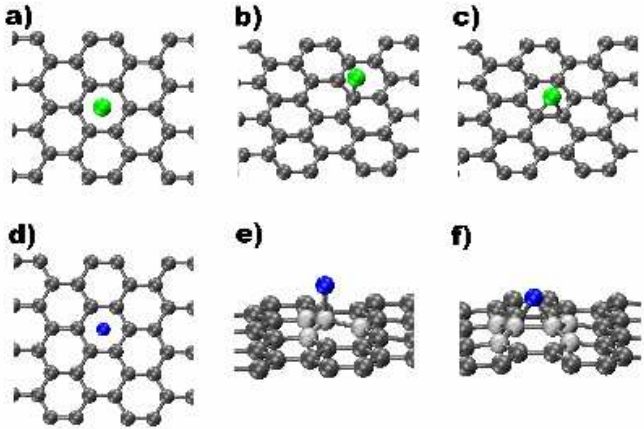}
\caption{(Color online) Schematic representations of the optimized geometries for the adsorption of B ({\it a, b, c} ) and N ({\it d, e, f}) on graphene. Figures
{\it a} and {\it d} are associated with the {\it hcp} site; {\it b} and {\it e} represent the {\it top} site while {\it c} and {\it f} stands for the {\it brigde} site. The clearer atoms in {\it e} and {\it f} indicate the carbon atoms pulled out the graphene plane.}
\label{sites}
\end{figure}
 
We first analyze the stability of a single B and a single N atom at the surface of a graphene layer. We 
studied three different local minima for each dopant: on {\it top} of a C atom; on a {\it bridge} site, in 
between two C atoms; and at the center of the hexagon ({\it hcp}). For both dopants, B and N, the most stable 
adsorption position is at the bridge site. Fig. \ref{sites} shows the possible locations and the relaxed 
configurations for  B and N. From a structural point of view, the adsorption of a B atom does not 
cause any significant distortion on the graphene sheet, while the N atom pulls two of the C atoms out of the 
plane. The environment around these C atoms becomes sp$^3$-like. This can be clearly 
observed in Fig. \ref{sites}e and \ref{sites}f.  The optimized structural parameters for these geometries and the energy difference for each site and each 
dopant, relative to most stable conformation is shown in Table \ref{tab01}. 
%%%%%%%%%%%%%%%%%%%%%%%%%%%%%%%%%%%%%%%%%%%%%%%%%%%%%%%%%%%%%%%%%%%%%
%\begin{table}[!htb]
%\caption{Variation of the energy for different adsorption sites relative  to the bridge site for B and N. All energies are in eV.}
%\label{tab01}
%\begin{tabular}{cccc}
%\hline
%\hline
%Dopant & top & hcp\\ 
%\hline
%B & 0.06 & 0.13\\
%N &  0.69 & 0.73\\
%\hline
%\hline
%\end{tabular}
%\end{table}
%%%%%%%%%%%%%%%%%%%%%%%%%%%%%%%%%%%%%%%%%%%%%%%%%%%%%%%%%%%%%%%%%%%%%

%%%%%%%%%%%%%%%%%%%%%%%%%%%%%%%%%%%%%%%%%%%%%%%%%%%%%%%%%%%%%%%%%%%%%
\begin{table}[!htb]
\caption{Variation of the energy ($\Delta$E) for different adsorption sites and  the distance between the dopant (D) and the closest carbon atom ($d_{D-C}$) for each site. All energies are in eV  and all distances are in \AA.}
\label{tab01}
\begin{tabular}{cccc}
\hline
\hline
Dopant & site & $\Delta$E & $d_{D-C}$\\ 
\hline
      &  top &  0.06 & 1.76\\
 B   &  hcp  & 0.13 &  2.24\\
      &  bridge& 0.00 & 1.86\\
 \hline
      & top & 0.73& 1.50\\
   N & hcp& 0.69 & 3.30\\
      & bridge & 0.00 & 1.49\\ 
\hline
\hline
\end{tabular}
\end{table}
%%%%%%%%%%%%%%%%%%%%%%%%%%%%%%%%%%%%%%%%%%%%%%%%%%%%%%%%%%%%%%%%%%%%%

\begin{figure}[ht]
\includegraphics[angle=0,width=9.0cm ]{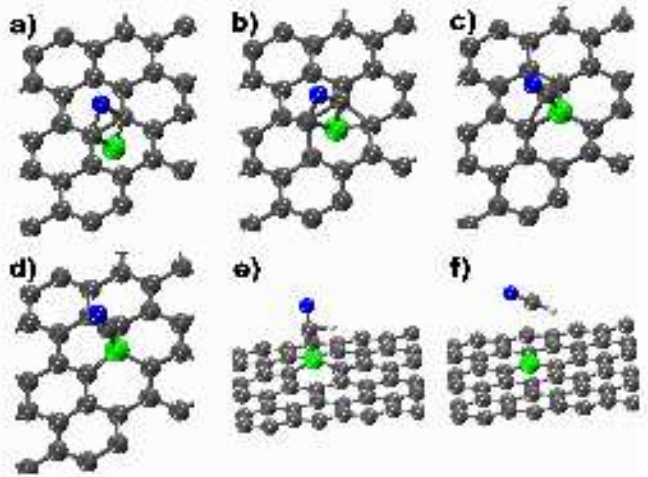}
\caption{(Color online) Representative geometries, along the structural optimization, for the atom substitution process ({\it a-d}) and for the radical removal ({\it e, f}). Larger, green spheres represent B, while smaller blue spheres N. C and H are grey and white respectively.}
\label{sites2}
\end{figure}

We proceed  to the analysis when the graphene membrane is exposed to two different environments with one type of atom in each side.
We expect that, when the graphene membrane is exposed to different mediums in each side, the atoms will
tend to be localized at the most stable position for each atom, as described above. As for both atoms the most stable site 
is the bridge site, we placed one B  atom in one side and one N atom on the other side of the 
membrane bridging theses carbon atoms. We then let the system relax to its equilibrium structure, where all the forces are minimized. The N atom binds to the layer, distorting it as shown before. The 
bonds between the C atom that is pulled out of the surface and its neighboring C atoms will be elongated
and consequently weakened. The B atom 
represents an exception to 
the octet rule\cite{19}, and it will be energetically more stable when bonded to three atoms. As the C-C bonds in the 
graphene membrane are weakened, the B atom will be pulled towards the surface, binding 
strongly to these three C atoms and replacing the original C atom from the membrane. 
Besides the energetic gain due to the three bonds that the B atom makes, another positive 
factor is related to size effects: B has an atomic radius comparable to C, such that there will 
not have much stress in the lattice due to the B incorporation. This mechanism has no activation barrier 
and is schematically shown in Fig. \ref{sites2}.

In the process described before, we were able to substitute a C atom by a B atom in the graphene layer, but 
there is still a C-N radical bonded to the dopant (Fig. \ref{sites2}d). The physical properties of this 
system will be different than a standard doping; the radical should be removed in order for the 
doping process to be accomplished. To achieve this, we insert H atoms into the system.  In our calculation, we have inserted an H atom nearby the C-N radical, as shown in Fig. \ref{sites2}e. The H atom will strongly bind to the C atom, forming HCN, which is 
spontaneously removed. The final result is a perfect doping of the graphene layer, as shown in 
Fig. \ref{sites2}f. As in the first part of the process, this mechanism does not have any activation barrier. 
We also tested the H atom connected with the N atom, however in this situation the radical was not removed.  
The energy difference between these two structures (1. H connected with the C atom of the radical {\it and} 
2. H attached to the N atom of radical) is $0.66 eV$ being the first situation energetically favorable.
The radical removal reaction can be described as:

 $Graphene_{BCN} + H  \rightarrow  Graphene_B + HCN$,\\
  where ($Graphene_{BCN}$) stands for Boron doped graphene coupled to a C-N radical plus a H atom,  creating Boron-doped graphene ($Graphene_B$) plus a free HCN. The reaction is exothermic, liberating an energy of 2.7 eV.

\begin{figure}[ht]
\includegraphics[angle=0,width=9.cm ]{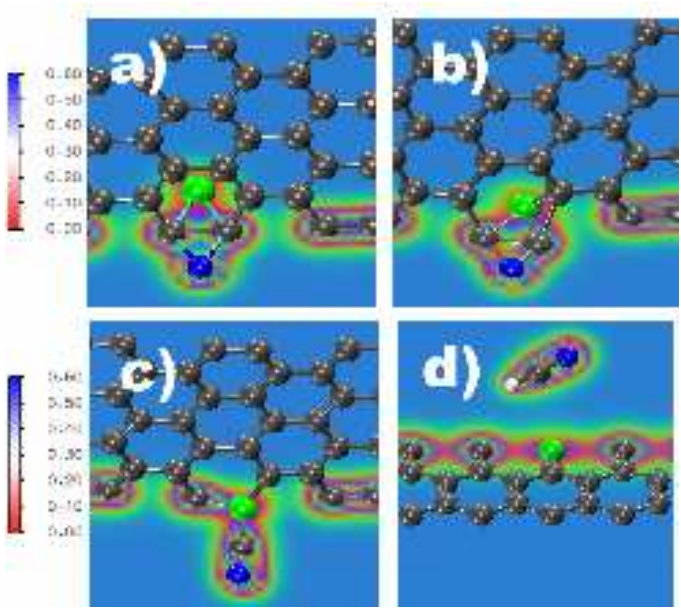}
\caption{(Color online) Total charge densities for different steps of the doping process. The scale is in e/\AA. }
\label{charge}
\end{figure}

The doping mechanism can be better understood by looking the
variation of the strength of each bond during the process. In Figure \ref{charge} we show a total 
charge density contour plot for different stages of the doping process. The red color in the scale 
bar indicates low densities and the blue color indicates high charge densities. 
The strong coupling between the N atom and graphene is clear in Fig. 
\ref{charge}a. In Fig \ref{charge}b we observe that the B atom is more connected with the right C coupled with N. In Fig. \ref{charge}c we present the plot for the last stage of the B incorporation and we can observe that the C-N bond is stronger than the B-C bond, this being part of the reason why this radical is removed when H is inserted. The B atom makes three bonds with the C atoms of the graphene lattice and the bond between the B and the carbon atom that is pushed out from the sheet is mediated by a weak lone pair shared by B and C atoms.

\begin{figure}[ht]
\includegraphics[angle=0,width=9.0cm ]{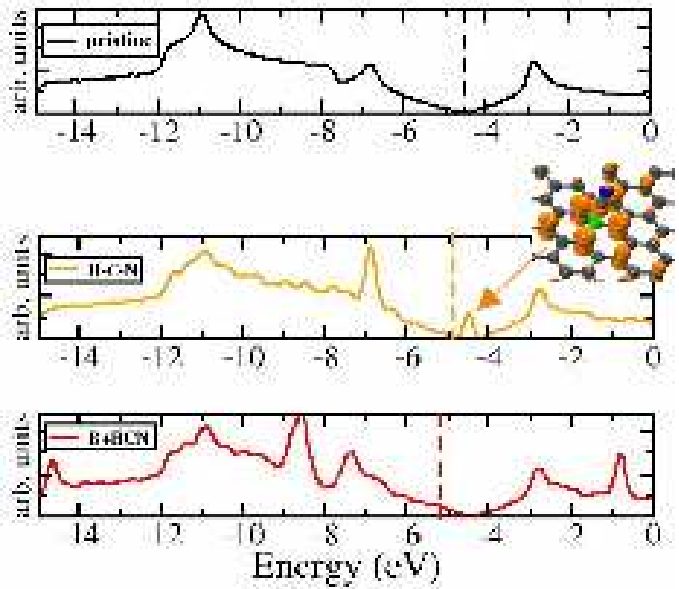}
\caption{(Color online) Total density of states for different doping stages. The graphs are aligned by the peak located at
-11eV. The vertical lines indicate the Fermi energy. The inset is a plot of the local charge density, in the (B-C-N) structure,  around the level created by the doping.}
\label{dos}
\end{figure}

In Figure \ref{dos} we present the density of states (DOS) of graphene (pristine), Boron doped graphene attached with the C-N radical (B-C-N) and Boron doped graphene with free HCN (B+HCN). For B-C-N we observe, above the Dirac point, the existence of an unoccupied level. This level, shown in the inset of Fig \ref{dos}, is originated from the orbital's re-hybridization due to Boron incorporation, and has major contributions from the carbon atoms surrounding the B atom, as well as contributions from the C and N present on the radical. When the H atom is inserted, this level is filled by the H's electron, and the radical will detach from the
sheet. The DOS of the final structure (B+HCN) is also presented, and we can see that the impurity  level has 
disappeared. Also, the Fermi energy shifts down with respect to the Dirac point (where the DOS is zero), indicating that Boron acts like a p-type dopant (acceptor),
increasing the DOS in the Fermi level for this doping concentration.

We also analyzed in more detail the structural and electronic properties for the final geometry of our process(Fig 2f), i.e, a substitutional B atom with a HCN molecule nearby. This is important since, in some cases, molecules adsorbed to graphene can 
also induce doping graphene\cite{12}. We obtained a B-C bond length of 1.52 \AA\  which is in good agreement with an experimental STM measurement of 1.59 \AA\  for the B-C bond length \cite{STM}. This result is also in agreement with  previous theoretical calculations\cite{Ferro}, where the B-C bond length is found to be 1.53 \AA. The nearest-neighbor C-C bonds are compressed by  0.015 \AA\ when compared to undoped  graphene. This clearly shows that the distortions are strongly localized around the B site. 

\begin{figure}[h]
\includegraphics[angle=0,width=9.0cm ]{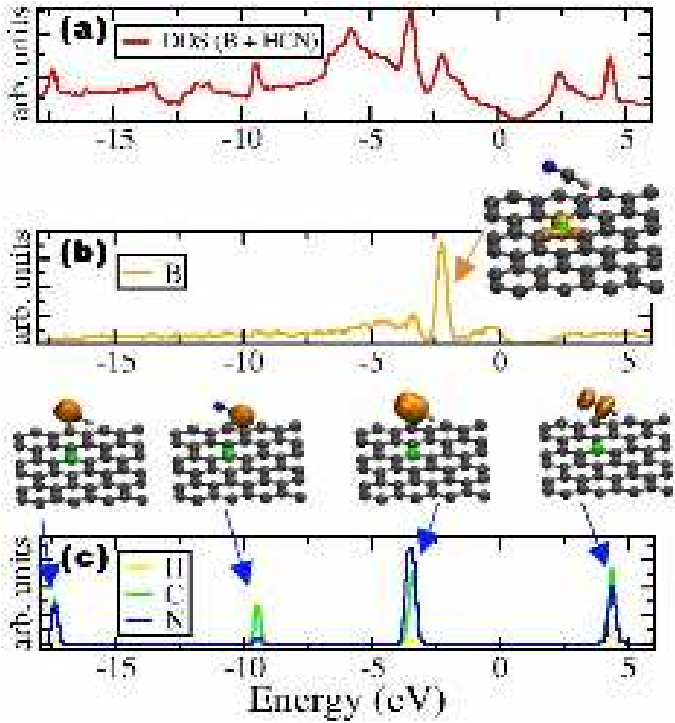}
\caption{(Color online)DOS (upper panel) and PDOS (lower panels) for the final structure presented in Fig 2f. The insets are local charge densities associated with the energy levels.  
The Fermi energy is at zero.  Because isolated levels correspond to delta functions in the DOS, for display purposes, we have artificially broadened the DOS by 0.1 eV. }
\label{dos1}
\end{figure}

For the final structure, Fig \ref{dos1} shows the decomposition of the total DOS. It is composed of discrete
HCN levels (lower panel),  and the doped graphene levels (middle panel). The local charge densities for these levels is presented on the insets, and 
we can clearly observe that HCN just weakly binds to graphene, and that the doped structure represents a real p-type doping of graphene.

The mechanism described here is not solely valid for the exact configuration described above. It also happens when the B
and the N atoms are located at {\it top} sites. More importantly, test calculations also showed that the B atom can be incorporated when the N atom is 
substituted by P, O, CH and NH, showing that our proposition is quite general and robust.

In conclusion, performing {\it ab initio} electronic structure calculations we have proposed an effective way to substitutionally dope graphene. 
This can be done by using graphene as a membrane to separate two different mediums. 
By this process, in principle, small atoms such as B could be readily incorporated through a barrierless process 
when the other side of the membrane is exposed to an environment that structurally distorts graphene.
The methodology described here should also be valid for doping other Carbon nanostructures such as
wide nanoribbons and  carbon nanotubes.

We would like to thank Prof. Ant\^onio J. R. da Silva and Wendel A. Alves for fruitful discussions. 
RBP thanks UFABC for financial support. GMD and AF were supported by Brazilian agencies FAPESP and CNPq.


\begin{thebibliography}{10}

\bibitem{1}
K. S. Novoselov, A. K. Geim, S. V. Morozov, D. Jiang, M. I. Katsnelson, V.
  Grigoreva, S. V. Dubonos, and A. A. Firsov, Nature {\bf 438}, 197 (2005).

\bibitem{2}
Y. Zhang, ; Y-W. Tan, H. L. Stormer, and P. Kim, Nature {\bf 438}, 201 (2005).

\bibitem{3}
S. Stankovich, D.A. Dikin, G.H.B. Dommett, K.M. Kohlhaas, E.J. Zimney, E.A.
  Stach, R.D. Piner, S.T. Nguyen, R.S. Ruoff, Nature {\bf 442} 282-286 (2006);.

\bibitem{4}
S. T. Nguyen, R. S. Ruoff, Nature {\bf 442}, 282 (2006).

\bibitem{5}
A. K. Geim, K. S. Novoselov, Nature Materials {\bf 6}, 183 (2007).

\bibitem{6}
X. Wang, Y. Ouyang, X. Li, H. Wang, J. Guo and H. Dai, Phys. Rev. Lett. {\bf
  100}, 206803 (2008).

\bibitem{7}
G. M. Dalpian, J. R. Chelikowsky, Phys. Rev. Lett. {\bf 96}, 226802 (2006).

\bibitem{8}
D. J. Norris, A. L. Efros, and S. C. Erwin, Science {\bf 319}, 1776 (2008).

\bibitem{9}
K. Nordlund, J. Keinonen, T. Mattila, Phys. Rev. Lett. {\bf 77}, 699 (1996).

\bibitem{10}
T. C. Mike Chung, Y. Jeong, Q. Chen, A. Kleinhammes, Y. Wu, J. Am. Chem. Soc.
  {\bf 130}, 6668 (2008).

\bibitem{12}
T. O. Wehling, K. S. Novoselov, S. V. Morozov, E. E. Vdovin, M. I. Katsnelson,
  A. K. Geim, and A. I. Lichtenstein, Nano Lett. {\bf 8}, 173 (2008).

\bibitem{11}
R. H. Miwa, T. B. Martins, and A. Fazzio, Nanotechnology, {\bf 19},
  155708(1)-155708(7) (2008).

\bibitem{13}
T. B. Martins, R. H. Miwa, A. J. R. da Silva and A. Fazzio, Phys. Rev. Lett.
  {\bf 98}, 196803 (2007).

\bibitem{14}
T. J. Booth, P. Blake, R. R. Nair, D. Jiang, E. W. Hill, U. Bangert, A.
  Bleloch, M. Gass, K. S. Novoselov, M. I. Katsnelson, A. K. Geim, Nano
  Letters, {\bf8} 2442 (2008).

\bibitem{bunch}
J. S. Bunch, S. S. Verbridge, J. S. Alden, A. M. van der Zande, J. M. Parpia,
  H. G. Craighead, and P. L. McEuen, Nano Letters, {\bf8} 2458 (2008).

\bibitem{15}
K. Capelle, Braz. J. Phys {\bf36} 1318 (2006). 

\bibitem{16}
J. P. Perdew, K. Burke and M. Ernzerhof, Phys. Rev. Lett. {\bf 77}, 3865
  (1996).

\bibitem{17}
N. Troullier, J. L. Martins, Phys. Rev. B. {\bf 43}, 1993 (1991).

\bibitem{18}
J. M. Soler, {\it et al}. J. Phys.: Condens. Matter {\bf 14}, 2745 (2002).

\bibitem{emilio}
E. Artacho, D. S\'anchez-Portal, P. Ordej\'on, A. Garcia, and J. M. Soler,
  Phys. Status Solidi B {\bf 215}, 809 (1990).

\bibitem{pack}
H. J. Monkhorst and J. D. Pack, Phys. Rev. B {\bf 13}, 5188 (1976).

\bibitem{19}
G. N. Lewis, Valence and the Structure of the Atoms and Molecules Ed. Dover
  :New York, (1966).

\bibitem{STM}
M. Endo, T. Hayshi, S-H. Hong, T. Enoki, and M. Dresselhaus, J. Appl. Phys.
  {\bf 90} 5670 (2001).

\bibitem{Ferro}
Y. Ferro, F. Marinelli, A. Allouche, and C. Brosset, J. Chem. Phys {\bf 118},
  5650 (2003).


\end{thebibliography}
\end{document}